# COMPETING TECHNOLOGIES:
# DISTURBANCE, SELECTION, AND THE POSSIBILITIES OF LOCK-IN

Loet Leydesdorff and Peter Van den Besselaar
University of Amsterdam, The Netherlands

## Abstract

Arthur's (1988) model for competing technologies is discussed from the perspective of evolution theory.  Using Arthur's own model for the simulation, we show that 'lock-ins' can be suppressed by adding reflexivity or uncertainty on the side of consumers.  Competing technologies then tend to remain in competition.  From an evolutionary perspective, lock-ins and prevailing equilibrium can be considered as different trajectories of the techno-economic systems under study.

Our simulation results suggest that technological developments which affect the natural preferences of consumers do not induce changes in trajectory, while changes in network parameters of a technology sometimes induce ordered substitution processes.  These substitution processes have been shown empirically (e.g., Fisher & Prey, 1971), but hitherto they have been insufficiently understood from the perspective of evolutionary modelling.  Implications for technology policies are discussed.

**Keywords:** competition, lock-in, alternative technologies, substitution, trajectory

## Arthur's (1988) model

Arthur generalized his model from so-called P\lya-urn models for the purpose of studying economic processes like standardization, network effects, and so-called 'increasing returns' (David, 1985).  As in the case of path-dependency in the probabilities of drawings in urn models, one can assume that how previous adopters have chosen from among competing technologies will matter for individual consumers (cf. Leydesdorff 1992 and 1995).

The two competing technologies are labeled $A$ and $B$.  These are cross-tabled with two types of agents, $R$ and $S$, with different 'natural inclinations' towards $A$ and $B$.  In *Table 1*, $a_R$ represents the natural inclination of $R$-agents towards type $A$ technology, and $b_R$ a (lower) inclination towards $B$.  Similarly, one can attribute parameters $a_S$ and $b_S$ to $S$-agents ($b_S > a_S$).

|          | Technology $A$ | Technology $B$ |
|----------|:--------------:|:--------------:|
| $R$-agent | $a_R + r\mathrm{n}_A$ | $b_R + r\mathrm{n}_B$ |
| $S$-agent | $a_S + s\mathrm{n}_A$ | $b_S + s\mathrm{n}_B$ |

**Table 1**. *Returns to adopting A or B, given $n_A$ and $n_B$ previous adopters of A and B.*
*(The model assumes that $a_R > b_R$ and that $b_S > a_S$. Both r and s are positive.)*

The network effects are modelled as independent terms, again differently for $R$-type agents and $S$-type agents.  The appeal of a technology is increased by previous adopters with a term $r$ (lower case) for each $R$-type agents, and $s$ for $S$-type agents.  If $R$-type and $S$-type agents arrive on the market randomly, the theory of random walks predicts that this competition will necessarily lock-in on either side ($A$ or $B$).

*Figure 1* shows the results of ten runs in a population of 10,000 adopters with parameter values 0.8 for $a_R$ and $b_S$, 0.2 for $a_S$ and $b_R$, and 0.01 for $r$ and $s$. The line in the middle corresponds to a 50% market share for each technology. As predicted, lock-in occurs in all cases, although not necessarily before the end of this simulation using 10,000 adopters.

The network effects are generated *endogenously* as a consequence of the values of parameters $r$ and $s$. If we reduce these two parameters with 50% to 0.005 in the above model, lock-in will often not occur in a population of this size (10,000). The absorbing barriers are not caused by (static) market conditions, but structural in a random walk with path-dependent feedback.

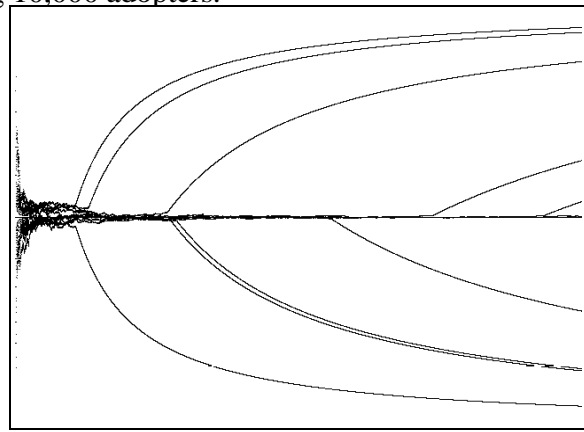

**Figure 1**  time
*Arthur's model as specified in Table 1; after 10 simulation runs*

## Technological leap-frogging

Is a technology that has lost out in the competition for lock-in able to leap-frog back in at a later date? Let us, for example, assume that a technological breakthrough is achieved in technology *B* after a lock-in in technology *A*. If the market is sufficiently large (($n_A + n_B$) > 2000), it can become attractive to the suppliers of technology *B* to invest in recapturing this market. The breakthrough is modelled in the next program given this market size and the condition that technology *A* has become dominant to the extent of capturing two-thirds of the market. What are the chances for technology *B*?

The technological breakthrough first operates on intrinsic inclinations because it changes the functional characteristics of the technology (for example, the price/performance ratio), and only upon diffusion can there be a network effect. These network effects will be discussed in a later section.

The breakthrough in technology *B* is simulated here by resetting the 'natural inclination' towards technology *B* for *R*-type adopters to 2.0 (instead of 0.2 before), while their 'natural inclination' for technology *A* is reduced to 0.08 (versus 0.8 before). Analogously, the natural inclination of *S*-type adopters towards

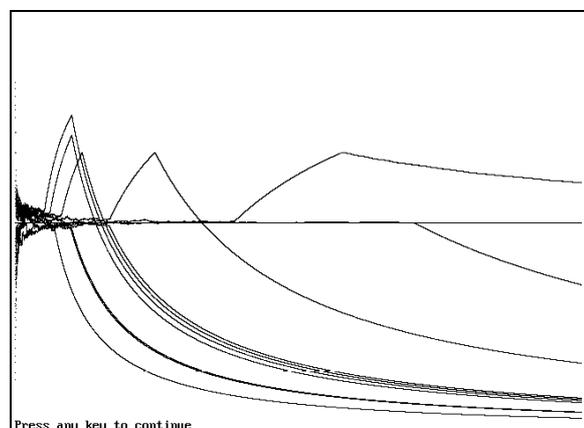

**Figure 2**  time
*Technological 'lock-out' and possible return to equilibrium (20,000 adopters)*

technology *A* is reduced with an order of magnitude to 0.02, and the inclination to the already preferred technology *B* is increased to 8.0. The lock-ins are *never* affected by these dramatic changes in the parameters. Results are similar to those in *Figure 1*. Thus, a lock-in prevents technological leap-frogging of a superior technology after the fact.

If we increase and decrease all relevant parameters with yet another order of magnitude, we force a reversal of the lock-in that we shall call a 'lock-out'. Interestingly enough, an in-between trajectory is visible when the market is sufficiently large. The system then returns to equilibrium instead of overshooting into a lock-in of technology *B* (*Figure 2*; based on 20,000 adopters).

## Uncertainty about relative market shares

In general, lock-in into technology *A* occurs when it has become more attractive for *S*-agents to buy this technology, despite their natural preference for technology *B*. From *Table 1*, we can learn that this is the case when:

$$a_S + s n_A > b_S + s n_B$$

thus:

$$s n_A - s n_B > b_S - a_S$$
$$(n_A - n_B) > (b_S - a_S)/s$$

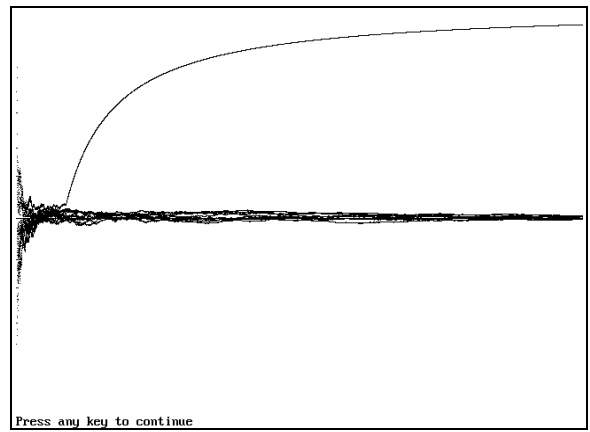

**Figure 3**        time 6
*Uncertainty about a difference smaller than 5% leads to suppression of the lock-in in nine out of ten cases*

In short: given the values for various parameters in the model, the lock-in depends only on the difference between the numbers of previous adopters $|n_A - n_B|$. When this difference surpasses a critical value, the system is locked-in. However, with increasing diffusion, the difference $(n_A - n_B)$ becomes smaller as a percentage of the market.

This relative decrease of the percentage has an economic interpretation. While consumers may be able to distinguish large differences in adoption between two technologies, decreasingly smaller percentages may become difficult to perceive. Let us, for example, assume that consumers can appreciate the difference in market penetration of technologies *A* and *B* as long as this difference is larger than 5% of the market. When the difference becomes smaller than 5%, adopters become uncertain, and we assume that they hold to their 'natural inclination' under this condition. *Figure 3* illustrates that lock-ins virtually disappear in this case.

## Reflexivity on the side of consumers

The Arthur-models assume that adopters are able to efficiently calculate their own net profits, and make decisions that follow market forces. However, people have a tendency to keep to their preferences even if they have to pay a price for them. In this section, we assume that switching to a technology other than the one 'naturally' prefered, can only be induced by the expectation of a net profit of 5% or larger. Thus, consumers will estimate their profits reflexively, and no longer react immediately to marginal profits.

The results of ten rounds of simulation under these conditions are similar to those in *Figure 3*. The lock-ins tend to disappear; they can occur incidentally as a consequence of the swings in market shares before equilibrium is achieved.

## 'Lock-out' because of changes in diffusion parameters

If, under the conditions specified above where technology *A* has captured two-thirds of a market with more than 2000 adopters, the network effect *r* (associated to *R*, and therefore with a preference for *A*) is reduced with three orders of magnitude, and the network effect *s* is increased with a factor thousand, the lock-ins are *not* affected. Thus, this system is robust against dramatic changes in diffusion effects.

However, if *s* vanishes at these conditions (*s* = 0) so that network effects disappear for *S*-type agents, equilibrium tends to be restored (*Figure 4*). This is an analytical consequence of the model specified in *Table 1*. Thus, after the fact of a lock-in one should invest in technologies that counter-act on network externalities rather than on technological breakthroughs that affect natural inclinations.

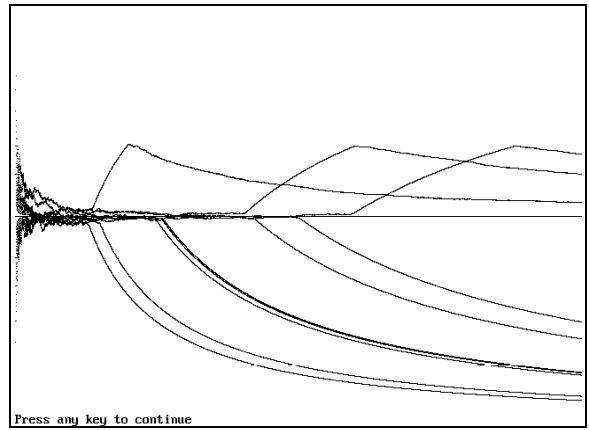

**Figure 4** time 6
*S-type agents do no longer profit from the network externalities (s = 0)*

## Locked-in versus Locking-in

These results raise the question whether network externalities should be attributed to choices by previous adopters or to the technologies involved. If we associate *r* (along the column in *Table 1*) with technology *A*, and correspondingly *s* with technology *B*, we obtain a model that is highly sensitive to changes in both *s* and *r*. For example, if the network effect *r* is reduced by only 50% (to 0.005) under the conditions above%where technology *A* has captured two-thirds of a market with more than 2000 adopters%the lock-ins for technology *A* instantaneously revert into lock-ins for technology *B*.

Interestingly enough, not only a decrease in *r* leads to this effect, but also an increase in *s*, that is, if the network effects of the non-dominant technology are gradually strengthened, for example, because of generational drift in the population. Thus, the lock-out in this case is an effect of changes in the network parameters of the technology which at that moment is no longer being traded. *S*-type agents (which arrive randomly) change their purchasing behaviour, and thereby force a change in lock-in.

Restoration of equilibrium can in this case not be achieved by changing the network parameters of technologies. One needs a third mechanism like uncertainty among adopters to induce a return to equilibrium. For example, if we combine the last simulation with the above condition of keeping to one's original preference when the difference in market share is less than 5% (given a market of 10,000 adopters),[1] the system obtains a window for returning to equilibrium. This is demonstrated in *Figure 5* for a population of 30,000.

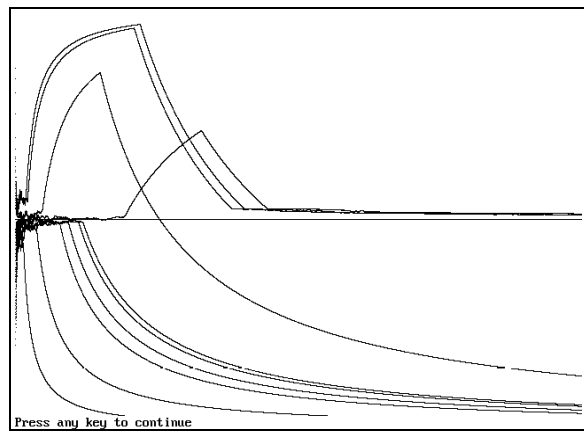

**Figure 5** time 6
*Lock-in, lock-out and return to equilibrium in a population of 30,000 adopters*

## Summary and Discussion

Lock-ins can be considered as discrete long-term expectations of network effects. Selections at the network level are a consequence of the relational factors in the variation. If one assumes that adopters have a reflexive capacity to estimate uncertainties, categorical lock-ins tend to disappear in the simulations in favour of a long-term trajectory of competitive equilibrium. A co-evolution between competing technologies may not be exceptional in view of the prevailing reflexivity and uncertainty at the actor level.

---

[1] This rise of the threshold in number of adopters for making uncertainty relevant is necessary, since otherwise almost no lock-ins occur (see *Figure 2*).

Our simulations show that trajectory transitions can be induced by changing network parameters. In this model the emergence of a new technology affecting the natural inclinations of consumers does not lead to a transition if there is already a lock-in. In their study of hyper-selection in innovation processes, Bruckner *et al.* (1994) showed that only within niches can the separatrix between the two basins of attraction sometimes be tunneled given a stochastic model. This conclusion accords with our results: if sufficiently present, network parameters become dominant (cf. Leydesdorff 1994).

Furthermore, we could show that infrastructural changes can sometimes lead to *ordered* substitution processes, notably following the lock-in line of the substituting technology. Although they have been demonstrated empirically (e.g., Fisher & Prey, 1971), substitution processes have been insufficiently understood from the perspective of evolutionary modelling. Not the emergence of a new technology, but structural adjustments in the dynamics of the network seem to determine the dissolution of one lock-in or another given a choice between competing technologies (cf. Freeman & Perez, 1988; David & Foray, 1994).